# Multi-Walled Nanotube Faceting Unravelled

I. Leven,[1,*] R. Guerra,[2,3,*] A. Vanossi,[2,3] E. Tosatti,[2,3,4] and O. Hod[1]

[1] *Department of Physical Chemistry, School of Chemistry, The Raymond and Beverly Sackler Faculty of Exact Sciences and The Sackler Center for Computational Molecular and Materials Science, Tel Aviv University, Tel Aviv 6997801, Israel*

[2] *International School for Advanced Studies (SISSA), Via Bonomea 265, 34136 Trieste, Italy*

[3] *CNR-IOM Democritos National Simulation Center, Via Bonomea 265, 34136 Trieste, Italy*

[4] *The Abdus Salam International Centre for Theoretical Physics (ICTP), Strada Costiera 11, 34151 Trieste, Italy*

*\* These authors contributed equally to this study.*

## Abstract

Nanotubes hold great promise for the miniaturization of advanced technologies. Their exceptional physical properties are intimately related to their detailed morphological and crystal structure. Importantly, circumferential faceting of multi-walled nanotubes acts to reinforce their mechanical strength and alter their tribological and electronic properties. Nevertheless, the full nature of this significant phenomenon remains to be explored. Here, nanotube faceting is fully rationalized in terms of interlayer registry patterns. We find that, regardless of the nanotube identity, faceting requires chiral angle matching between adjacent layers. Above a critical diameter that corresponds well with experimental findings, achiral multi-walled nanotubes display evenly spaced extended axial facets, whose number is determined by the interlayer difference in circumferential unit cells. Elongated helical facets, most commonly observed in experiments, appear in nanotubes exhibiting a small interlayer chiral angle mismatch. In the case of uncorrelated wall chiralities faceting is suppressed, and outer layer corrugation induced by the Moiré superlattice is obtained in excellent agreement with experiment. Based on this we offer an explanation to the relative abundance of faceting found in multi-walled boron nitride nanotubes with respect to their carbon nanotube counterparts.

## Introduction:

As their name suggests, nanotubes are traditionally thought of as cylindrical structures possessing a circular cross section. Nevertheless, under certain conditions multi-walled nanotubes (MWNTs) exhibit circumferential faceting resulting in polygonal cross sections.[1-8] This, in turn, leads to considerable reinforcement of their mechanical properties[7,9] thus paving the way for fulfilling their potential as next generation ultra-high strength materials.[10-12] Furthermore, it may have considerable impact on their tribological, electronic, and optical properties. A thorough understanding of the nature and origin of this phenomenon is thus crucial for the rational design of robust and durable nano-electro-mechanical devices that can withstand repeated mechanical load.

The underlying mechanism for nanotube faceting clearly involves a balance between inter-layer attractive interactions gained at the faceted regions and mechanical strain accumulated at the apexes. This understanding has been the basis for several theoretical studies adopting the continuum model approach.[13-16] Such models provide valuable insights regarding the general phenomenon. However, they are limited in their ability to depict important system-specific characteristics that require a detailed atomistic description. This is clearly demonstrated by the work of Palser, who used a fully atomistic dispersion corrected anisotropic tight-binding Hamiltonian model for graphitic systems to study pentagonal faceting in nested multi-walled zigzag nanotubes.[17] Nevertheless, many important questions such as: what are the detailed conditions required for faceting to occur? What dictates the number of facets formed? And why is this phenomenon more commonly observed in multi-walled boron nitride nanotubes (MWBNNTs) than in their carbon counterparts? remain open.

In the present study, we identify the atomistic origin of nanotube faceting resulting from extended inter-layer registry patterns that appear between the curved hexagonal lattices forming the nanotube walls. We find that when two adjacent walls have matching chiral angles, their curvature difference forms localized out-of-registry regions that are evenly spaced along the circumference of the tube and are separated by extended in-registry arcs. Upon structural relaxation the former become apexes whereas the latter form commensurate facets. Interestingly, the number of facets is dictated by the difference in circumferential unit cells between two adjacent layers. Furthermore, the facets can be either axial or spiral depending on the chiral angle difference of adjacent layers. Finally, the critical tube diameter required for faceting is found to be between 5-13 nm depending on tube chirality, in good agreement with experimental observations.[7] Based on these findings we offer an explanation for the relative abundance of faceting in inorganic nanotubes with respect to their graphitic counterparts.

**Achiral Double-Walled Nanotubes:**

Let us first discuss achiral double walled nanotubes (DWNTs). In this case, four different types of systems exists where the inner and outer layers can be either zigzag (ZZ) or armchair (AC). In Fig. 1 we present relaxed structures of ZZ@ZZ, AC@AC, and ZZ@AC double-walled carbon nanotubes (DWCNTs) and double-walled boron nitride nanotubes (DWBNNTs) with outer diameter in the range of 5-20 nm. Here, the notation ZZ@AC, for example, represents a DWNT with an inner ZZ tube inside an outer AC shell. The most prominent feature evident in the figure is that for achiral DWNTs cross-sectional polygonalization occurs when the inner and outer shells share the same chiral angle whereas for the mixed ZZ@AC systems the cross section remains circular. Furthermore, both the carbon and BN systems studied exhibit a critical diameter of 5-7 nm and 9-13 nm for the AC@AC- and ZZ@ZZ-DWNTs, respectively, beyond which faceting appears. This is in remarkable agreement with recent experimental observations suggesting a critical faceting diameter of ~12 nm in MWBNNTs.[7]

In order to elucidate the interlayer effects underlying the formation of facets we visualize in Fig. 1 the degree of local interlayer commensurability by coloring each atom of the outer shell according to the value of its local registry index (LRI, see methods section and supporting information for a detailed explanation). Focusing first on the pre-optimized circular cross section structures (see upper row of Fig. 1) a unique pattern is revealed for the ZZ@ZZ and AC@AC DWNTs where localized out-of-registry areas (marked in red) appear along the circular circumference and are separated by better registry (white and blue regions) arcs. Upon structural relaxation the former develop into vertices whereas the latter form the facets resulting in an overall improvement of the interlayer registry (see movie in the supporting information). Furthermore, the vertices that bear most of the mechanical tension are characterized by larger interlayer registry mismatch as compared to the facets that form practically optimally stacked flat interfaces. Interestingly, along the circumference of the pre-optimized ZZ@AC DWNTs studied a uniform interlayer registry picture appears. Hence, in the lack of a symmetry breaking driving force acting to promote polygonization the systems remain circular after optimization and no faceting occurs. A similar picture is found for AC@ZZ systems (see supporting information).

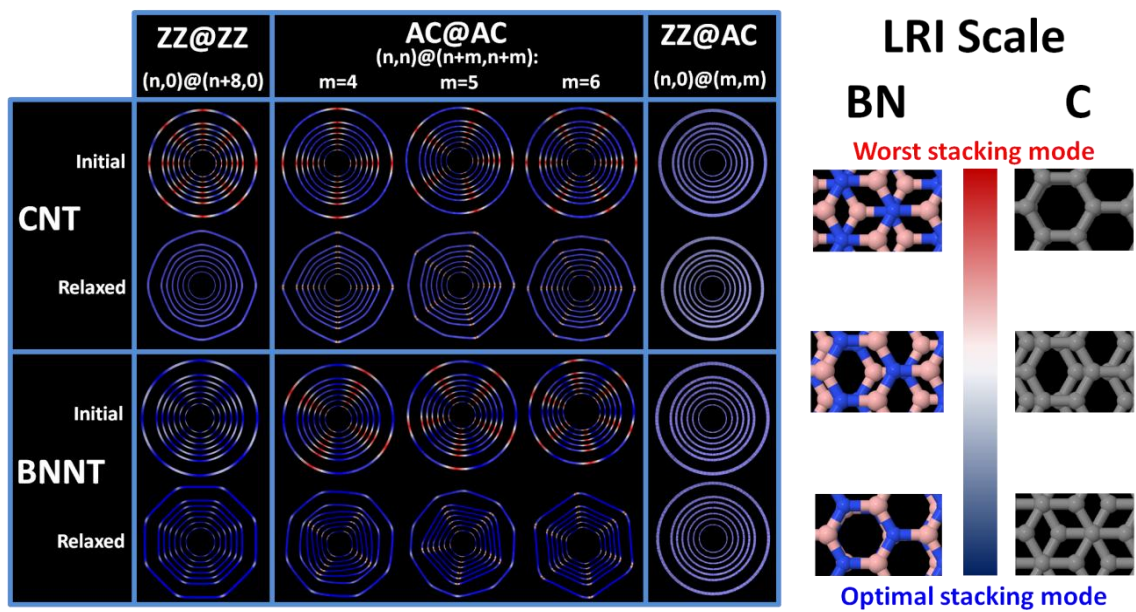

**Figure 1: Relaxed achiral DWNTs geometries and LRI patterns.** Schematic representation of achiral DWCNTs (upper rows) and DWBNNTs (lower rows) showing their structure and local registry patterns before and after geometry relaxation. The pre-optimized structures represent the lowest energy circular cross section interlayer configuration of each system. Optimization has been performed using the Airebo intra-layer and Kolmogorov-Crespi inter-layer potentials for the DWCNTs and the Tersoff intra-layer and *h*-BN ILP inter-layer potentials for the DWBNNTs (see methods section). Each frame includes seven DWNTs with diameters in the range of 5-20 nm. Five groups of DWNTs are presented (from left to right): ZZ@ZZ (n,0)@(n+8,0); AC@AC (n,n)@(n+4,n+4), (n,n)@(n+5,n+5), and (n,n)@(n+6,n+6); and ZZ@AC. For the ZZ@ZZ systems we choose n=55, 80, 105, 130, 155, 180, and 243; For the AC@AC systems we use n=31, 46, 60, 75, 89, 104, and 140; and for the ZZ@AC systems we consider the following set: (54,0)@(36,36), (80,0)@(51,51), (104,0)@(65,65), (130,0)@(80,80), (154,0)@(94,94), (179,0)@(108,108), (241,0)@(144,144). The LRI color bar appearing on the right ranges from blue to red representing the optimal and worst staking modes of graphene and hexagonal boron nitride, respectively. Grey, pink, and blue spheres represent carbon, boron, and nitrogen atoms, respectively. These colors should not be confused with the LRI patterns coloring depicted for all DWNTs on the left panel.

An important question that remains unanswered is what dictates the number of facets formed in the AC@AC and ZZ@ZZ systems? Answering this question requires the understanding of what determines how many incommensurate regions appear along the circumference of the pre-optimized circular DWNTs. This turns out to be a purely geometric problem that can be readily addressed using the LRI method. To demonstrate this we have considered three sets of AC@AC DWNTs with varying interlayer distance, including the (n,n)@(n+4,n+4), (n,n)@(n+5,n+5), and (n,n)@(n+6,n+6) systems, where the notation (n,n)@(n+4,n+4), for example, represents a DWNT with an inner (n,n) shell inside and outer (n+4,n+4) wall (see Fig. 1). Comparing the pre-optimized registry patterns of the three sets it becomes evident that the number of localized incommensurate regions (and hence the eventual number of vertices found in the optimized structures) is given by the difference in the number of circumferential unit-cells between the inner and outer walls regardless of the DWNT diameter. Namely, for the (n,n)@(n+4,n+4) system four incommensurate regions form and, similarly, for the (n,n)@(n+5,n+5) and (n,n)@(n+6,n+6) systems five and six incommensurate regions form, respectively, for all values of the nanotube index, n, considered. Each arc connecting two localized incommensurate regions is a circumferential completion of one unit-cell misfit between the inner and outer layers. Interestingly, the mismatch in number of circumferential unit-cells tends to evenly spread along the circumference of the circular DWNTs even when the walls do not have a mutual rotational symmetry. However, when the inner and outer tube indices share a greatest common divisor (GCD) larger than one, the DWNT obtains a GCD-fold rotational symmetry and the registry patterns become fully periodic.[18,19] We note that upon structural relaxation of all AC@AC DWCNTs studied and some of the larger AC@AC DWBNNTs considered secondary kinks appear between each pair of main vertices thus doubling the number of facets formed.

**Chiral Double-Walled Nanotubes:**

We now turn to address the more general case of DWNTs consisting of, at least, one chiral layer. Similar to their achiral counterparts, monochiral DWBNNTs,[20] whose inner and outer walls share the same chiral angle ($\Delta\theta = \theta_{Outer} - \theta_{Inner} = 0°$), present registry patterns with axial symmetry (see upper left panel of Fig. 2). As before, localized incommensurate regions appear along the circumference of the pre-optimized structures. Nevertheless, the (near-)rotational symmetry obtained for the achiral systems is lost and their degree of registry mismatch varies. Here, as well, upon structural relaxation the incommensurate regions thin down and form corners that are separated by axial facets with increased interlayer commensurability (middle left panel). This is more clearly seen in the interlayer distance analysis (lower left panel) where the facets present a nearly optimal (~3.33 Å) interlayer separation whereas the vertices show an increase of up to 0.2 Å.

The four right columns of Fig. 2 show various bichiral DWBNNTs, whose inner and outer walls differ in their chiral angles ($\Delta\theta \neq 0°$). These systems present non-axial registry patterns spiraling around the circumference of the tube with a helical angle that depends on the chiral angle difference between the two walls. For the bichiral DWBNNTs of smaller chiral angles mismatch considered ($\Delta\theta = 0.253°$ and $0.657°$) long incommensurate stripes appear separated by spiraling nearly-commensurate arcs (upper panels of the second and third columns). Similar to the mono-chiral DWBNNT case, upon structural relaxation the incommensurate stripes become thinner and form corners with increased interlayer distance whereas the nearly commensurate arcs turn into helical facets with close to optimal interlayer spacing (see middle and lower panels). This indeed is the most common faceting pattern observed in experiment.[2,3,5,6] For larger chiral angles mismatch (two rightmost columns) the incommensurate regions become discontinuous (upper panels) resulting in non-facetted optimized structures (middle and lower panels).

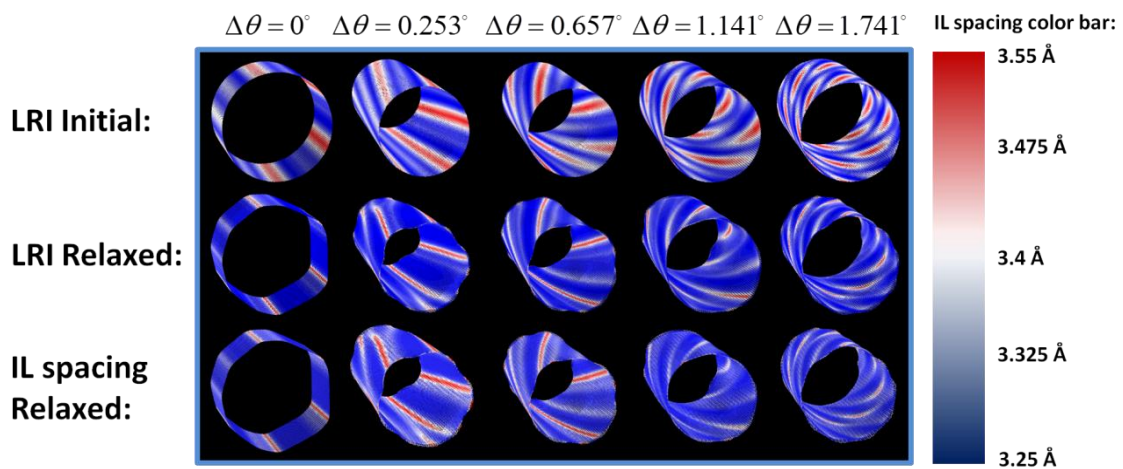

**Figure 2: Relaxed chiral DWBNNTs geometries, LRI patterns, and interlayer distance.** Schematic representation of (120,100)@(126,105) (leftmost column), (60,60)@(66,65) (second column), (70,70)@(77,74) (third column), (68,68)@(75,70) (fourth column), and (71,71)@(80,72) (rightmost column) DWBNNTs showing their local registry

patterns before (top row) and after (middle row) geometry relaxation performed using the Tersoff intra-layer and *h*-BN ILP inter-layer potentials (see methods section). The color bar on the right refers to the interlayer spacing of the different systems presented in the bottom row. The LRI color bar is the same as in Fig. 1. The chiral angle difference, $\Delta\theta$, between the inner and outer shells is indicated above the respective columns.

## Two-Dimensional Mapping:

This complex behavior can be fully rationalized by the theory of Moiré patterns in planar mismatched hexagonal lattices.[21] To this end, the registry patterns appearing in a DWNT with circular cross section are mapped onto the corresponding flat bilayer system. This is achieved by first unrolling the tube shells to obtain two infinite parallel ribbons followed by a contraction of the wider ribbon, representing the outer shell, to match the width of the inner ribbon thus mimicking curvature effects (see Fig. 3a).[18,21] The resulting planar bilayer is thus periodic along the circumferential direction. The obtained Moiré patterns form a super structure whose lattice vectors are given by:[22]

$$\begin{cases} \boldsymbol{L}_1^M = \frac{\sqrt{3}d\sin(\bar{\theta})}{2(1+c_h'/c_h)\sin\left(\frac{\Delta\theta}{2}\right)}\{2\hat{\mathbf{x}} + [(1-c_h'/c_h)\cot\left(\frac{\Delta\theta}{2}\right) - (1+c_h'/c_h)\cot(\bar{\theta})]\hat{\mathbf{y}}\} \\ \boldsymbol{L}_2^M = \frac{\sqrt{3}d\sin(\bar{\theta}-\frac{\pi}{3})}{2(1+c_h'/c_h)\sin\left(\frac{\Delta\theta}{2}\right)}\{2\hat{\mathbf{x}} + [(1-c_h'/c_h)\cot\left(\frac{\Delta\theta}{2}\right) - (1+c_h'/c_h)\cot(\bar{\theta}-\frac{\pi}{3})]\hat{\mathbf{y}}\}, \end{cases} \quad (1)$$

where $c_h$ and $c_h'$ are the chiral vectors lengths of the inner and outer tube walls, respectively, $\bar{\theta}$ is the average chiral angle of the two tube walls, $\Delta\theta$ is the corresponding chiral angle difference, $d$ is the nearest-neighbour inter-atomic distance within each tube wall, and $\{\hat{\mathbf{x}},\hat{\mathbf{y}}\}$ is the orthogonal reference frame defined by two unit vectors lying along the chiral and translation vectors directions of the inner tube wall.

For both achiral and monochiral nanotubes $\Delta\theta = 0°$ and the Moiré lattice vectors become parallel to the translational vector of the tube with diverging length thus resulting in the extended axial LRI patterns appearing both in the curved (Fig. 1 and left column of Fig. 2) and in the planar (Fig. 3c,d) representation of the double walled nanotubes. Upon increasing $\Delta\theta$ the Moiré lattice vectors deviate from the translational vector thus developing the commonly observed spiraling patterns along the circumference of the tube (Fig. 2 and 3e). Importantly, since the length of the Moiré lattice vectors reduces with increasing chiral angle difference (see Fig. 3b) the density of lattice mismatch regions increases (see right column of Fig. 2 and Fig. 3f) thus suppressing the formation of extended facets. This clearly demonstrates that faceting requires interlayer chiral angle matching. We note, however, that even in the case of $\Delta\theta = 1.741°$, with lattice vector length of $|L_1^M| =$ 4.305 nm and $|L_2^M| =$ 17.157 nm, the optimized structure exhibits interlayer distance variations that correlate with the registry patterns. Notably, similar variations have been recently observed on the surface of carbon nanotubes.[23,24]

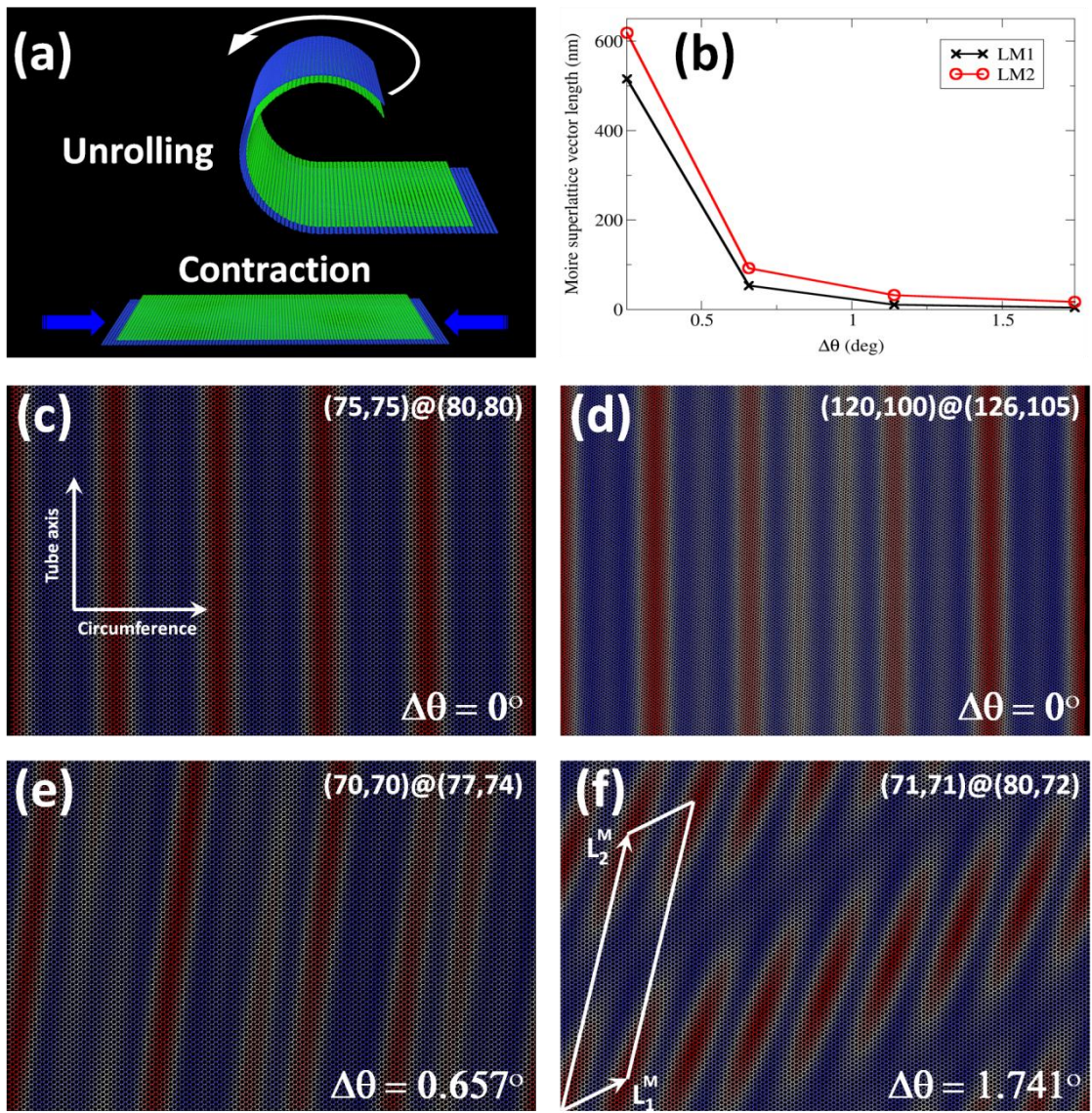

**Figure 3: Planar mapping of DWNT registry pattern.** (a) Schematic illustration of NT unrolling and compression resulting in the planar mapping of the DWNT registry patterns. The atoms of the inner and outer walls are colored in green and blue, respectively. (b) Dependence of the Moiré lattice vectors length on the chiral angle

difference between the DWNT walls as calculated using Eq. (1). (c),(d) Planar mapping of the registry patterns of the (75,75)@(80,80) and (120,100)@(126,105) DWBNNTs with $\Delta\theta=0^o$ characterized by axial registry patterns. (e),(f) Planar mapping of the helical registry patterns of the (70,70)@(77,74) and (71,71)@(80,72) DWBNNTs with $\Delta\theta=0.657^o$ and $\Delta\theta=1.741^o$, respectively. The Moiré superlattice cell of the (71,71)@(80,72) DWBNNT calculated using Eq. (1) is represented by the white parallelogram in panel (f). The horizontal and vertical axes in panels (c)-(f) correspond to the circumferential and axial directions of the DWBNNTs, respectively. The LRI color bar is the same as in Fig. 1.

## Summary and Conclusions

To summarize, the origin of multi-walled nanotube circumferential faceting is found to be a direct consequence of interlayer lattice registry patterns. The appearance of extended facets requires chiral angle matching between adjacent nanotube layers. For two achiral neighboring layers of the same type (armchair or zigzag) extended axial facets with (nearly-)perfect rotational symmetry are formed, the number of which is dictated by the corresponding difference in the number of circumferential unit cells. The critical diameter for faceting in these systems is found to be 5-13 nm, in good agreement with experimental findings. Similarly, mono-chiral layers present extended axial facets but with reduced rotational symmetry. Bi-chiral adjacent layers exhibit helical facets whose length decreases and helix angle increases with increasing interlayer chiral angle difference up to a point where facet formation is suppressed resulting in a corrugated nanotube surface, in good agreement with experimental findings. This, in turn, provides an explanation to why faceting is more abundant in MWBNNTs than in MWCNTs. The polar nature of the hetero-nuclear BN covalent bonds in MWBNNTs introduces interlayer electrostatic interactions[25] that are sufficient to induce inter-wall chiral angle correlation[5,26-28] required for the formation of facets. The lack of such interactions in MWCNTs that are composed of non-polar homo-nuclear CC covalent bonds often[29,30,31] results in practically random chiral angle distribution of the different nanotube layers,[23,24,26,32-38] which may explain why mostly non-facetted structures appear. It is therefore evident that gaining control over their interlayer registry matching provides a route for the mechanical enforcement[7] as well as tribological, electronic, and thermal properties tuning of MWNTs.[39]

## Methods

Geometry optimization of all DWNTs considered have been performed using quenched dynamics techniques with dedicated intra- and inter-layer classical force-fields. For CNTs the intra-layer interactions have been described utilizing both the Tersoff[40] potential (see supporting information) and the reactive Airebo[41] force-field adopting the parameterization of Lindsay and Broido.[42] The inter-layer interactions of these systems have been described by the registry dependent Kolmogorov-Crespi[43] potential. For the intra-layer interactions of BNNTS we have used the Tersoff[40] force-field as parameterized by Sevkin et al. for BN based systems[44] along with our recently developed $h$-BN inter-layer potential ($h$-BN ILP) with fixed partial charges (see supporting information).[45] Cyclic boundary conditions have been applied for all achiral and monochiral (bearing the same chirality) DWNTs. We further verify the robustness of our results toward potential many-body dispersion forces screening effects that have been neglected in the parameterization procedure of the $h$-BN ILP (see supporting information).

For the registry analysis we have extended the global registry index (GRI) method, which quantifies the interlayer stacking registry in rigid layered materials,[18,46-48] by defining the local registry index that characterizes the local degree of lattice commensurability in various regions along the circumference of the nanotube. In short, in the original GRI approach a single parameter is calculated as sums and differences of all projected atomic-centered circle overlaps between adjacent layers to characterize the overall registry matching of the system at a given interlayer configuration. In the LRI approach each atomic center is assigned a number that indicates the local degree of registry in its immediate surrounding environment. This is achieved by calculating the projected overlaps between circles (or Gaussians) assigned to a given atom and one of its nearest neighbors in one layer and all circles (Gaussians) of its adjacent layer using the same system-dependent circle radii (Gaussian standard deviation), formula, and normalization as in the GRI approach. This procedure is repeated for all nearest neighbors of the given atom and the average result is assigned to this atom as its LRI such that a value of 0 marks good local registry (lowest interlayer energy) and 1 stands for bad local registry (highest interlayer energy). The LRI map is then obtained by plotting one of the NT layers and using a color scheme in which an atom with the highest LRI value of 1 is colored in red and an atom having the lowest LRI value of 0 is colored in blue (see Figs. 1-3). For further details see supporting information.

## Acknowledgements

O.H. acknowledges the Lise-Meitner Minerva Center for Computational Quantum Chemistry and the Center for Nanoscience and Nanotechnology at Tel-Aviv University for their generous financial support. Work in Trieste was carried out under ERC Grant 320796 MODPHYSFRICT. EU COST Action MP1303 is also gratefully acknowledged.